\documentclass[sn-aps]{sn-jnl}% Style for submissions to Nature Portfolio journals

\usepackage{microtype}%
\usepackage{graphicx}%
\usepackage{multirow}%
\usepackage{amsmath,amssymb,amsfonts}%
\usepackage{amsthm}%
\usepackage{mathrsfs}%
\usepackage[title]{appendix}%
\usepackage{xcolor}%
\usepackage{textcomp}%
\usepackage{manyfoot}%
\usepackage{booktabs}%
\usepackage{algorithm}%
\usepackage{algorithmicx}%
\usepackage{algpseudocode}%
\usepackage{listings}%
\usepackage{siunitx}%
\usepackage{xspace}
\xspaceaddexceptions{)}
\usepackage[acronyms]{glossaries}

\newacronym{hdc}{HDC}{Hyperdimensional Computing}

\newcommand{\hdc}{\gls{hdc}\xspace}

\newcommand{\tcam}{TCAM\xspace}
\newcommand{\sram}{SRAM\xspace}
\newcommand{\finfet}{FinFET\xspace}

\newcommand{\fesquid}{FeSQUID\xspace}

\newcommand{\nbits}{\ensuremath{n_{bits}}\xspace}
\newcommand{\nmatch}{\ensuremath{n_{match}}\xspace}
\newcommand{\vml}{\ensuremath{V_{ML}}\xspace}%
\hypersetup{
    colorlinks = false,
    hidelinks = true
}
\usepackage{subcaption}
\usepackage{gnuplottex}
\usepackage{xparse}
\ExplSyntaxOn
\DeclareExpandableDocumentCommand{\convertlen}{ O{cm} m }
{
	\dim_to_decimal_in_unit:nn { #2 } { 1 #1 } cm
}
\ExplSyntaxOff
\usepackage{cleveref}%
\crefname{enumi}{Step}{Steps}
\crefname{section}{Sec.}{Sec.}
\crefname{subsection}{Sec.}{Sec.}
\crefname{figure}{Fig.}{Fig.}
\crefname{algocf}{Algorithm}{Algorithms}
\crefname{algorithm}{Algorithm}{Algorithms}
\crefname{algocf}{Algorithm}{Algorithms}
\crefname{table}{Tab.}{Tab.}
\crefname{equation}{Eq.}{Eq.}
\crefname{eqnarray}{Eq.}{Eq.}
\crefname{appendix}{Sec.}{Sec.}
\def\nobreakbefore{%
  \relax\ifvmode\else
    \ifhmode
      \ifdim\lastskip > 0pt\relax
        \unskip\nobreakspace
      \fi
    \fi
  \fi
}
\let\oldcite\cite
\renewcommand\cite{\nobreakbefore\oldcite}

\theoremstyle{thmstyleone}%
\theoremstyle{thmstyletwo}%

\theoremstyle{thmstylethree}%

\begin{document}

\title[Article Title]{Energy-Efficient Cryogenic Ternary Content Addressable Memory using Ferroelectric SQUID}


\author[1]{\fnm{Shamiul} \sur{Alam}}\email{salam10@vols.utk.edu}

\author[2]{\fnm{Simon} \sur{Thomann}}\email{s.thomann@tum.de}

\author[3]{\fnm{Shivendra Singh} \sur{Parihar}}\email{sshiven@iitk.ac.in}

\author[3]{\fnm{Yogesh Singh} \sur{Chauhan}}\email{chauhan@iitk.ac.in}

\author[4]{\fnm{Kai} \sur{Ni}}\email{kni@nd.edu}

\author[2]{\fnm{Hussam} \sur{Amrouch}}\email{amrouch@tum.de}

\author[1]{\fnm{Ahmedullah} \sur{Aziz}}\email{aziz@utk.edu}

\affil[1]{\orgdiv{Department of Electrcial Engineering and Computer Science}, \orgname{University of Tennessee}, \orgaddress{ \city{Knoxville}, \postcode{37996}, \state{TN}, \country{USA}}}

\affil[2]{\orgdiv{TUM School of Computation, Information and Technology}, \orgname{Technical University of Munich}, \orgaddress{\city{Munich}, \postcode{80992}, \country{Germany}}}

\affil[3]{\orgdiv{Department of Electrical Engineering}, \orgname{Indian Institute of Technology (IIT)}, \orgaddress{\city{Kanpur}, \postcode{208016}, \state{U.P.}, \country{India}}}

\affil[4]{\orgdiv{Department of Electrical Engineering}, \orgname{University of Notre Dame}, \orgaddress{\city{Notre Dame}, \postcode{46556}, \state{IN}, \country{USA}}}


\abstract{Ternary content addressable memories (TCAMs) are useful for certain computing tasks since they allow us to compare a search query with a whole dataset stored in the memory array. They can also unlock unique advantages for cryogenic applications like quantum computing, high-performance computing, and space exploration by improving speed and energy efficiency through parallel searching. This paper explores the design and implementation of a cryogenic ternary content addressable memory based on ferroelectric superconducting quantum interference devices (FeSQUIDs). The use of FeSQUID for designing the TCAM provides several unique advantages. First, we can get binary decisions (zero or non-zero voltage) for matching and mismatching conditions without using any peripheral circuitry. Moreover, the proposed TCAM needs ultra-low energy (1.36 aJ and 26.5 aJ average energy consumption for 1-bit binary and ternary search, respectively), thanks to the use of energy-efficient SQUIDs. Finally, we show the efficiency of FeSQUID through the brain-inspired application of Hyperdimensional Computing (HDC). Here, the FeSQUID-based TCAM implements the associative memory to support the highly parallel search needed in the inference step. We estimate an energy consumption of 89.4 fJ per vector comparison using a vector size of 10,000 bits. We also compare the FeSQUID-based TCAM array with the 5 nm FinFET-based cryogenic SRAM-based TCAM array and observe that the proposed FeSQUID-based TCAM array consumes over one order of magnitude lower energy while performing the same task.}


\maketitle

\section{Introduction}\label{sec1}

Cryogenic computing systems, capable of operating at/below 4 Kelvin temperature, have garnered renewed interest in recent years primarily due to their promise as control processor and memory in large-scale quantum computing systems \cite{alam2023cryogenic,hornibrook2015cryogenic,alam2022cryocim}. In addition, they are uniquely suited for exa-scale high-performance computing systems and space applications \cite{holmes2013energy,huang2022survey,das2017large}. A suitable cryogenic controller and memory system can facilitate the scaling of quantum computing systems up to thousands of qubits, by solving several existing challenges, including (i) the requirement of a large number of wires and interconnects to connect the qubits with currently used room temperature controller and memory, (ii) the possibility of a large amount of thermal noise propagation from room temperature to noise sensitive qubits, and (iii) the heat and noise generation by the lossy wires and interconnects \cite{hornibrook2015cryogenic, alam2023cryogenic}. Moreover, cryogenic systems based on superconducting devices (such as Josephson junctions and superconducting quantum interference devices (SQUID)) provide unparalleled speed (hundreds of gigahertz) and energy efficiency (sub-atto-joule switching energy) \cite{holmes2013energy}. This can be extremely useful for developing energy-efficient high-performance computing systems and space electronics. 

\begin{figure*}
\begin{center}
\includegraphics[width=\textwidth]{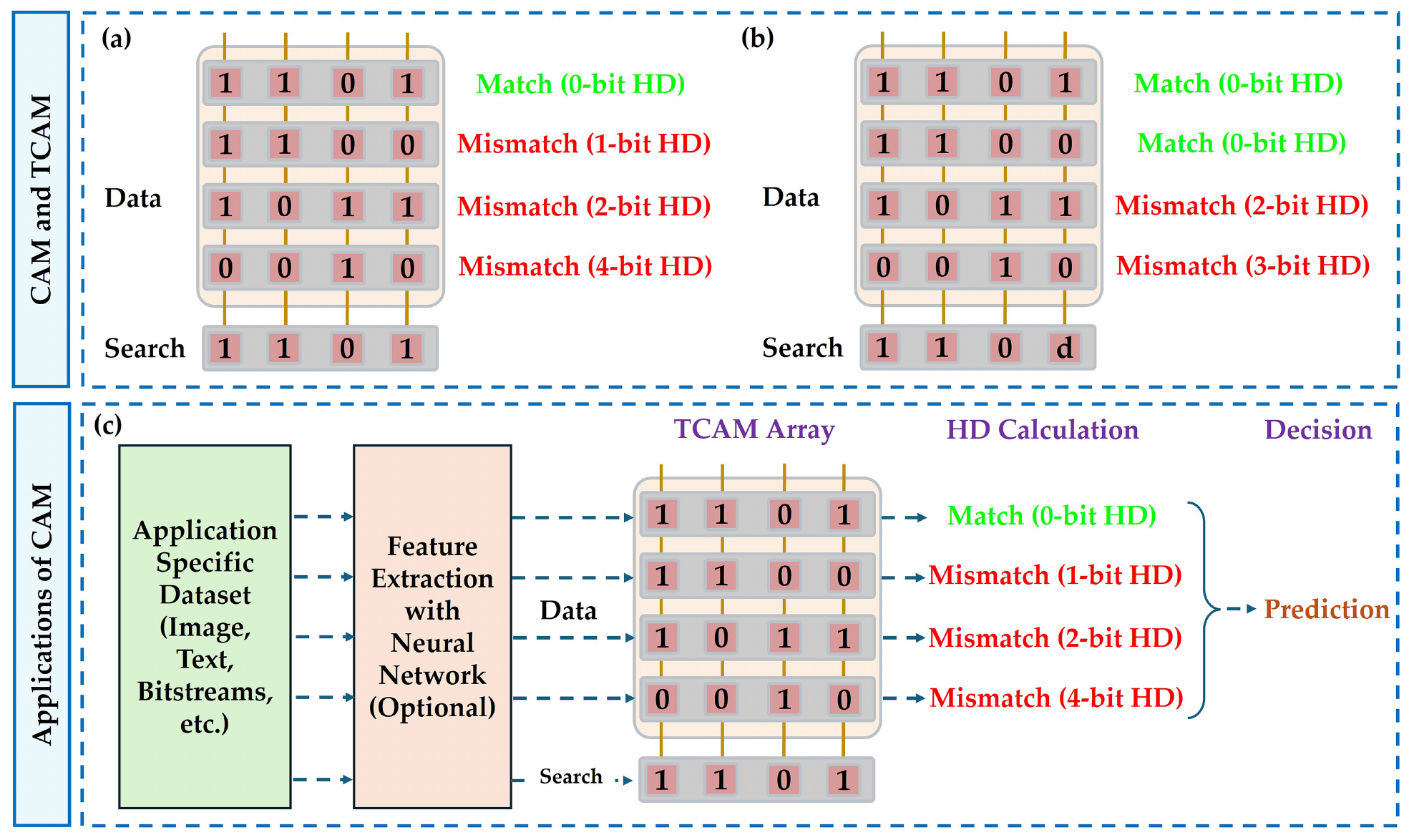}
\caption{Introduction to content addressable memory. Illustrations of the search mechanism with a (a) binary and (b) ternary CAM. (c) Block diagram showing the steps how a CAM can be used for AI applications such as classification and pattern recognition tasks.}
\label{fig:Intro}
\end{center}
\end{figure*}

One of the critical challenges in cryogenic systems is the need for high-speed, low-power memory solutions that can complement the computing infrastructure \cite{alam2023cryogenic}. In this quest, superconducting, non-superconducting, and hybrid technologies have been explored \cite{alam2023cryogenic,alam2021non, ghoshal1993superconductor,alam2022cryogenic, alam2021cryogenic,tannu2017cryogenic}. However, all of these technologies have unique challenges, and therefore, developing a suitable and scalable memory system for cryogenic applications has remained an ongoing pursuit. Content addressable memory (CAM) and ternary content addressable memory (TCAM) are specialized storage systems that can accelerate certain computing tasks by introducing the capability of comparing input data against stored data \cite{islam2023quantum, ni2019ferroelectric}. CAM performs searches based on binary data ('0' and '1') returning exact match/mismatch output, which TCAM allows an additional state (don't care, 'd') for the input data (Fig.~\ref{fig:Intro}(a and b)). The ternary nature of TCAMs allow for more flexible searches, which is useful in several applications including pattern recognition, network routing, and database management. Additionally, CAM and TCAM can be empowered to calculate the Hamming distance (HD) between the input search and stored data. This capability allows us to identify nearby matches (the lowest HD implies the closest match) along with the exact matching (HD = 0). CAMs and TCAMs with HD calculation capability can be powerful tools in error detection and correction for both classical and quantum domains, approximate searching, and other data-intensive applications. Additionally, TCAMs have recently been used to accelerate classification tasks, where the TCAM stores the features extracted from the dataset (images, texts, etc.) as the stored data and can provide the search result whenever the features for any search condition are applied as the input data (Fig.~\ref{fig:Intro}(c)) \cite{ni2019ferroelectric}. Based on the search results, TCAM can predict the classification result. Here, the parallel search capability of TCAMs becomes extremely useful for improving the speed and efficiency of the classification tasks. 

TCAMs can also be useful for cryogenic applications like quantum computing by efficiently handling quantum error correction decoding, high-performance computing by improving latency and energy efficiency, artificial intelligence (AI) by accelerating pattern recognition and classification, space exploration by assisting in data storage and retrieval, and decision-making algorithms, and data centers by improving large-scale search operations like database queries and search engines. This paper demonstrated the design and implementation of a cryogenic TCAM with exact search and HD calculation capabilities. Here, we utilize the ferroelectric SQUID (FeSQUID)-based cryogenic memory system that combines the ultra-high speed and energy efficiency of SQUIDs with the voltage-controlled non-volatility of ferroelectric materials. In the proposed TCAM, there are two modes- one for exact searching and another for HD calculation. Due to the use of FeSQUID, we get several advantages from the proposed TCAM when compared with the existing designs (both room temperature and cryogenic), including (i) the exact searching mode provides binary decisions for matching and mismatching without needing any peripheral circuitry, thanks to the superconducting behavior of SQUIDs, and (ii) extremely energy-efficient TCAM operation (1.36 aJ and 26.5 aJ average energy consumption for 1-bit CAM and TCAm search. respectively). Finally, we demonstrate the efficiency and advantages of our proposed TCAM by performing Hyperdimensional Computing (HDC) algorithm-based language recognition task. By implementing the associative memory that supports a highly parallelized search operation for the inference step with \fesquid-based \tcam arrays, a single 10,000-bit vector comparison consumes just 89.4 fJ. We also demonstrate that the FeSQUID-based TCAM consumes over one order of magnitude less energy compared to the cryogenic 5 nm FinFET SRAM-based TCAM to perform the same operation.

\section{Devices of Interest}\label{sec2}

In our proposed TCAM, we use FeSQUIDs as storage elements and heater cryotrons (hTrons) as the access devices to allow write and read operation of any specific FeSQUID in a large array scenario. In this section, we introduce these two unique devices. 

\subsection{Ferroelectric SQUID (FeSQUID)}\label{subsec21}

\begin{figure*}
\begin{center}
\includegraphics[width=\textwidth]{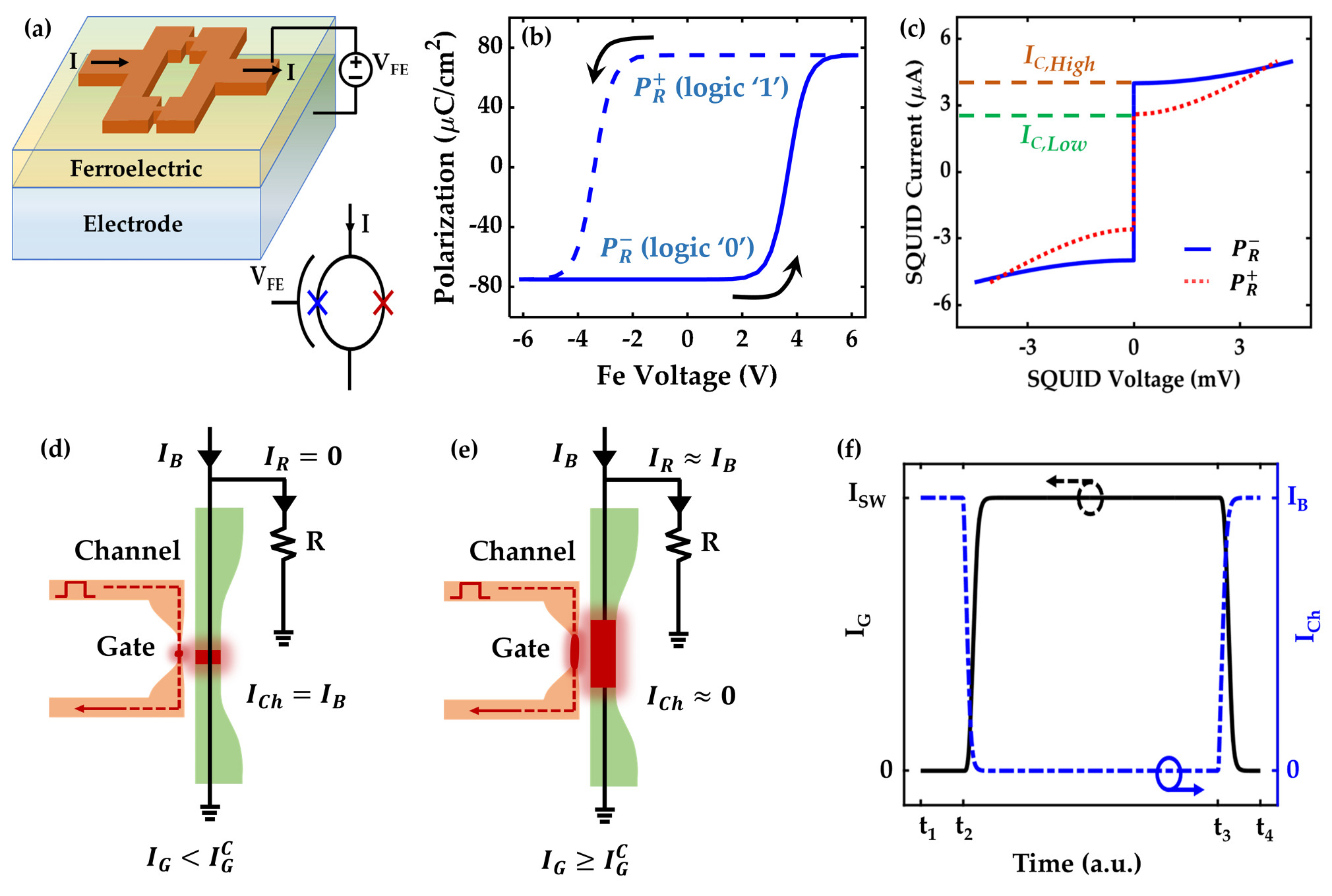}
\caption{(a) Device structure and circuit symbol of a FeSQUID. (b) Polarization-voltage characteristics of a Led Zirconium Titanate (PZT) ferroelectric material. (c) Current-voltage characteristics of the SQUID for two different polarization states of the ferroelectric material. Device structure and illustration of gate-controlled switching of a hTron channel when (d) $I_G < I_G^C$, keeping the channel in its superconducting state and (e) $I_G > I_G^C$, switching the channel to its resistive state. (f) Gate current-controlled switching of the hTron channel.}
\label{fig:Devices}
\end{center}
\end{figure*}

Despite extensive research over the last few decades, a robust method for voltage-controlled gating of SQUIDs remained elusive until the introduction of FeSQUID in 2021 \cite{suleiman2021nonvolatile}. In FeSQUID, a ferroelectric material is employed to modulate the superconducting behavior of the SQUID which also allows voltage control over the superconductivity. The proposed FeSQUID had a SQUID built with two parallel weak links on top of a ferroelectric material, the structure is shown in Fig.~\ref{fig:Devices}(a). Ferroelectric materials exhibit non-volatile polarization switching that can be controlled by an external voltage bias. Fig.~\ref{fig:Devices}(b) illustrates the voltage-controlled switching of the ferroelectric polarization ($P_{FE}$) of a led zirconium titanate (PZT) material at cryogenic temperatures. The internal $P_{FE}$ of a ferroelectric material generates surface charges that, in turn, induce electric fields, effectively injecting direct charge into the system \cite{huang2021direct}. Now, When a SQUID is fabricated on top of a ferroelectric layer, the superconducting material screens the charge bound at the interface. The amount of bound charge at the interface directly depends on the remnant polarization ($P_R$) of the ferroelectric. Specifically, negative remnant polarization ($P_{R}^-$) increases the surface-bound charge, while positive remnant polarization ($P_{R}^+$) decreases it \cite{suleiman2021nonvolatile,crassous2011nanoscale}. This alteration in surface charge modifies the carrier density, which in turn affects the critical temperature ($T_{C}$) of the superconductor and, consequently, the superconducting energy gap ($\Delta$). The relationship between $\Delta$ and $T_{C}$ can be described using Bardeen–Cooper–Schrieffer (BCS) theory \cite{bardeen1957theory,alam2020compact}:

\begin{equation} \Delta (T) = 1.763 k_B T_C ~\tanh\left(2.2\sqrt{\frac{T_C}{T} -1}\right) \end{equation}

where $T$ is the temperature and $k_B$ is the Boltzmann constant. The Ambegaokar-Baratoff (AB) theory \cite{ambegaokar1963tunneling} further explains how $\Delta(T)$ affects the critical current ($I_C$):

\begin{equation} I_C = \frac{\pi \Delta}{2q_e R_N} ~\tanh\left(\frac{\Delta}{2k_B T}\right) \end{equation}

Here, $q_e$ is the electron charge, and $R_N$ is the normal state resistance of the SQUID. Due to the non-volatile polarization states of the ferroelectric, two distinct levels of $I_C$ are observed in the SQUID's I-V characteristics. Specifically, the $P_{R}^-$ state of the ferroelectric results in a higher critical current ($I_{C,high}$), whereas the $P_{R}^+$ state leads to a lower critical current ($I_{C,low}$), as shown in Fig.~\ref{fig:Devices}(c). This unique voltage-controlled superconductivity of FeSQUID has been leveraged to implement a scalable memory system \cite{alam2022cryogenic}, a voltage-controlled Boolean logic family \cite{alam2024voltage}, and an in-memory computing system \cite{alam2023Dac} for cryogenic applications. 

\subsection{Heater Cryotron (hTron)} \label{subsec22}

To overcome the limitations of two-terminal Josephson junctions, three-terminal cryotron-based devices were developed, offering gate current-controlled switching of the superconducting channel between its superconducting and resistive states \cite{mccaughan2014superconducting}. A notable member of this family is hTron \cite{mccaughan2019superconducting}, which is capable of driving high impedances ($>100 ~k\Omega$) and supporting a large number of fan-outs due to its highly resistive state. Unlike Josephson junctions, hTron devices do not rely on superconducting loops, making them free from issues like flux trapping and scalability challenges \cite{mccaughan2014superconducting}. hTrons have already been used as an access device in different cryogenic memories \cite{alam2021cryogenic, nguyen2020cryogenic, alam2022cryogenic}, as an interface between superconductors and semiconductors \cite{mccaughan2019superconducting}, to design logic circuits \cite{alam2023reconfigurable,mccaughan2014superconducting, alam2024voltage, alam2024ultra}, in cryogenic neuromorphic systems \cite{islam2023review, islam2022dynamically, islam2023synapse}, and so on. 

The hTron is a four-terminal superconducting device driven by current. Two terminals serve as the gate, while the other two form the superconducting channel, as depicted in Fig.~\ref{fig:Devices}(d). The gate and channel are separated by a dielectric spacer that thermally connects but electrically isolates the two. When no gate current ($I_G$) is applied, the channel remains superconducting, assuming a given channel bias current ($I_{B}$). However, once $I_G$ is applied, the gate becomes resistive, generating thermal phonons that propagate through the dielectric spacer to the superconducting channel. Until $I_G$ increases beyond a certain threshold, the superconductivity of the channel persists \cite{mccaughan2014superconducting}. This behavior is shown in Fig.~\ref{fig:Devices}(d).

When the gate current ($I_G$) exceeds a critical threshold ($I_G^C$), enough thermal phonons with sufficient energy ($>2\delta$) are generated and transported to the channel. These phonons become able to disrupt the cooper pairs and suppress the superconductivity of the channel superconductor. Therefore, the channel's critical current ($I_{Ch}^C$) falls below the applied channel bias current ($I_{B}$). This causes the channel to transition into a high-impedance resistive state, redirecting the channel current ($I_{B}$) to the external circuit, as illustrated in Fig.~\ref{fig:Devices}(e). Fig.~\ref{fig:Devices}(f) shows the gate current-controlled switching of the channel between its superconducting and resistive states.

\section{Cryogenic TCAM based on FeSQUID and hTron}\label{sec3}

\begin{figure*}
\begin{center}
\includegraphics[width=0.85\textwidth]{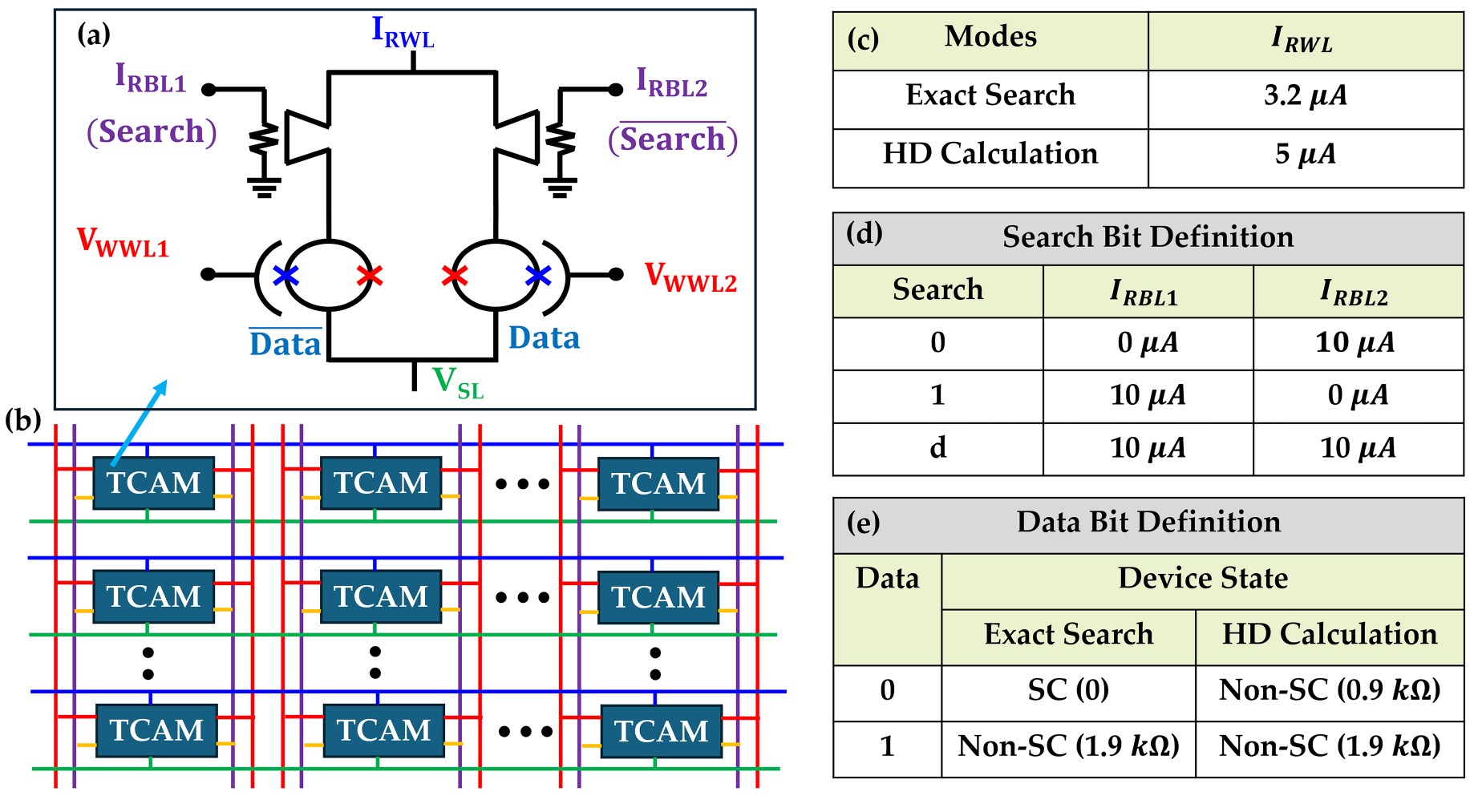}
\caption{The proposed cryogenic TCAM based on FeSQUID. (a) Circuit schematic of the proposed TCAM cell. (b) Illustration of the array-level organization of the proposed TCAM cells, where data will be stored in a row. (c) Values of RWL currents for two modes of the proposed TCAM. Definitions of the representation of (d) input search data in terms of RBL currents and (e) stored data in terms of FeSQUID's device states. }
\label{fig:CAM}
\end{center}
\end{figure*}

Fig.~\ref{fig:CAM}(a) shows the schematic of the proposed cryogenic TCAM. In this section, we discuss the design and working principles of this TCAM. 

\subsection{Design Principle} \label{subsec31}

The proposed TCAM is a modified version of FeSQUID-based memory cell, demonstrated in \cite{alam2022cryogenic}. In the FeSQUID-based memory cell, to store one bit of data, one FeSQUID and one hTron are connected in series and multiple cells are connected in parallel along each row of the memory array. However, in this work, we use two memory cells connected in parallel to each other to design a 1-bit TCAM. The operation of this TCAM cell is controlled by one read word line (RWL), two read bit lines (RBLs), two write word lines (WWLs), and one sense line (SL). The operating principle of the TCAM cell is discussed in detail in Section~\ref{subsec32}. Now, to build a large-scale TCAM, these TCAM cells need to be connected in parallel along the row where the shared RWL in each row will act as the match line (ML) to exhibit the matching/mismatching result (Fig.~\ref{fig:CAM}(b)). 

The proposed TCAM has two modes- one for exact search (binary result) and another for HD calculation (analog result). The mode of the TCAM can be selected by choosing a suitable bias current for the RWL of each TCAM. For exact search mode, the value of RWL current ($I_{RWL}$) needs to be chosen in a way so that we get superconducting and resistive states from SQUIDS for '0' and '1' data, respectively. According to Fig.~\ref{fig:Devices}(c), the range for this $I_{RWL}$ is -- $I_{C,Low}<I_{RWL}<I_{C,High}$. Due to the use of the superconducting state of SQUID and the parallel connection of all the TCAMs in a row (as shown in Fig.~\ref{fig:CAM}(b)), if there is any SQUID with a superconducting state, the ML of that row will have zero voltage. This provides us with the binary decision from the ML- either zero or some non-zero voltage for any amount of mismatch or complete matching, respectively. Now, for HD calculation mode, we do not want binary decisions for matching and mismatching, rather we want analog decisions with different levels of ML voltage for different levels of matching between the input and stored data. Therefore, we need to avoid the superconducting state of SQUIDs and that's where the use of FeSQUIDs becomes extremely useful because FeSQUIDs can show either superconducting/resistive states or only resistive state with two different resistance values for two states. In the HD calculation mode, we choose the value of $I_{RWL}$ in a way so that FeSQUIDs show two different resistance values for '0' and '1' states. According to Fig.~\ref{fig:Devices}(c), the required range of $I_{RWL}$ for this mode is - $I_{RWL}>I_{C,High}$. Fig.~\ref{fig:CAM}(c) shows the values of $I_{RWL}$ used for two TCAM modes in this work.

The FeSQUIDs are utilized to store the data as the ferroelectric polarization (similar to the memory operation) for the TCAM operations. One FeSQUID stores the data ($Data$) while the other stores the inverted version of data($\Bar{Data}$). To store $Data$ and $\Bar{data}$, we need to apply suitable voltages across the ferroelectric of FeSQUIDs through the WWLs ($V_{WWL1}$ and $V_{WWL2}$). Fig.~\ref{fig:CAM}(e) shows the definition of $Data$ and the corresponding device states. The input search data is applied through a combination of two gate currents (RBL currents, $I_{RBL1}$ and $I_{RBL2}$) of the two hTrons, which determine the switching of the hTrons. Fig.~\ref{fig:CAM}(d) shows the values of RBL currents for different input search data ($Search$ and $\Bar{Search}$).

\subsection{Wroking Principle} \label{subsec32}

Any TCAM search can be divided into two major operations- (i) storing the data inside the memory through memory write operation and (ii) searching any input data against the stored data through memory read operation. To write data into a FeSQUID-based memory cell, we apply appropriate voltage biases (positive or negative) across the ferroelectric layer. Specifically, we manipulate the polarization state of the ferroelectric material within the targeted FeSQUID cell by employing a V/2 biasing scheme. The appropriate voltage biases ($\pm V_{WRITE}$ or $\pm V_{WRITE}/2$) are applied to the write word lines (WWLs) and source lines (SLs), such that only the selected memory cell experiences the full $\pm V_{WRITE}$ across its ferroelectric layer. In this configuration, half-selected cells located in the same row or column as the targeted cell will experience $\pm V_{WRITE}/2$, while unselected cells, not sharing the same row or column, will experience 0 V. To ensure the write operation affects only the selected cell, the write voltage must be carefully chosen to satisfy the condition $\frac{1}{2}|V_{WRITE}| < |V_C| < |V_{WRITE}|$, where $V_C$ is the coercive voltage of the ferroelectric. 

As illustrated in Fig.~\ref{fig:Devices}(c), applying a current within the range $I_{C,Low} < I < I_{C,High}$ to the SQUID will result in either a superconducting (0 V) state for $P_R^-$ or a resistive (nonzero voltage) state for $P_R^+$, depending on the polarization state of the ferroelectric. This property is exploited for the exact search mode by applying a suitable current bias through the SQUID. First, all WWLs and SLs are grounded. Moreover, as seen in Fig.~\ref{fig:Devices}(c), if a current larger than $I_{C,High}$ is applied, the SQUID will show two different resistance values (and hence, two voltages) depending on the polarization states. This characteristic of SQUID is used for the HD calculation mode. RBL currents are responsible for determining the switching of hTrons connected in serial with each FeSQUID. The search data are represented with the RBL currents. 

\begin{figure*}
\begin{center}
\includegraphics[width=\textwidth]{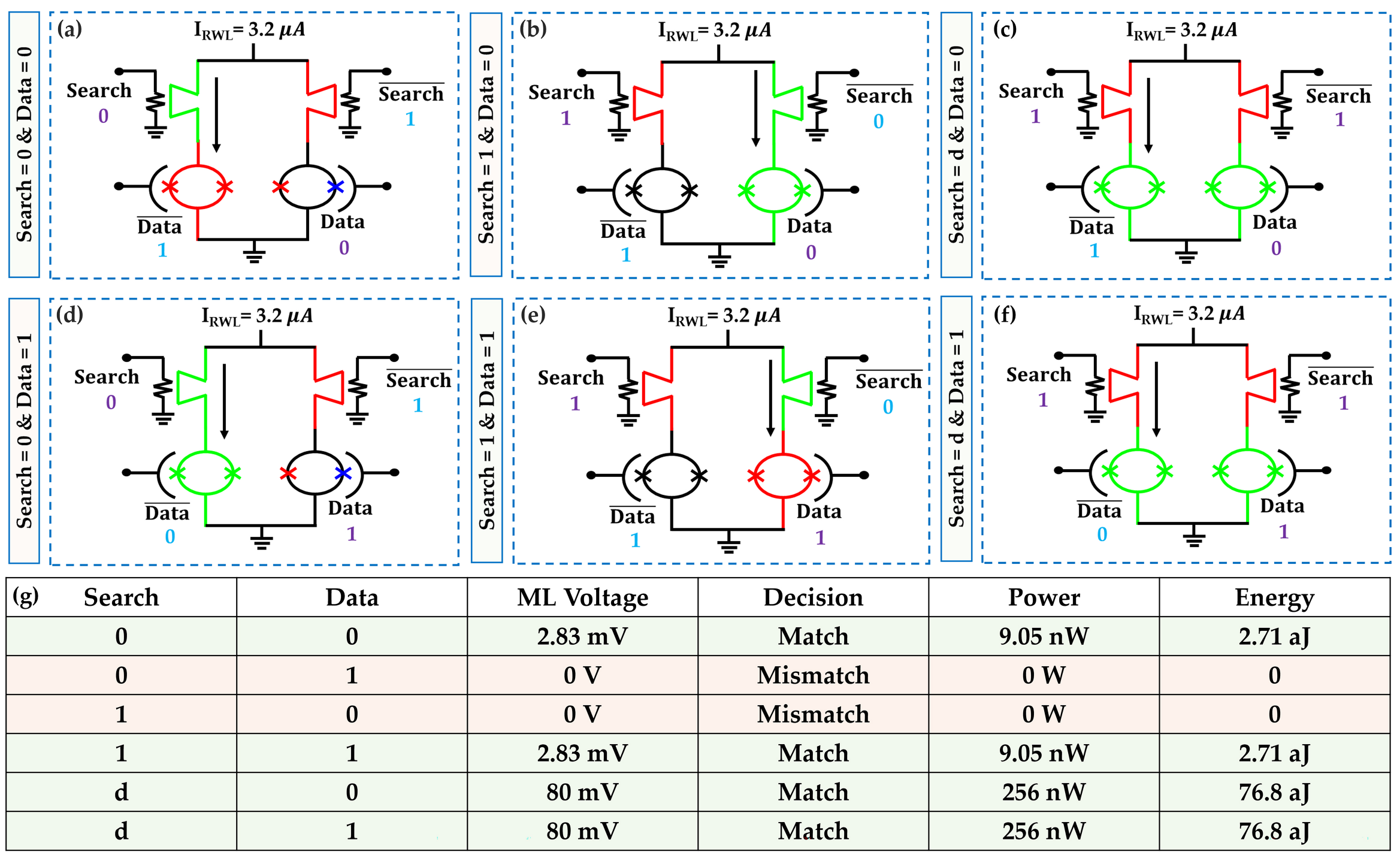}
\caption{Working Principle of the proposed TCAM in the exact search mode. Illustration of the working principle of the TCAM when (a) $Data = 0$ and $Search = 0$, (b)$Data = 0$ and $Search = 1$, (c) $Data = 0$ and $Search = d$, (d) $Data = 1$ and $Search = 0$, (e) $Data = 1$ and $Search = 1$,  and (f) $Data = 1$ and $Search = d$. (g) ML voltage, power, and energy consumption for 1-bit exact search with the proposed TCAM. }
\label{fig:ExactTCAM}
\end{center}
\end{figure*}

Based on the application of RWL and RBL currents, the ML voltage will depend on whether the input search matches or mismatches the stored data. Fig.~\ref{fig:ExactTCAM} shows the working principle of the exact search model of the proposed TCAM. Here, first, $Data$ and $\Bar{Data}$ are stored in the two FeSQUIDs with the help of WWLs. Then, when $Search$ and $\Bar{Search}$ are applied through the RBL currents (chosen according to Fig.~\ref{fig:CAM}(d)), and the suitable current is applied to the RWL (chosen according to Fig.~\ref{fig:CAM}(c)), one of the hTrons become superconducting while the other remains in a highly resistive state ($50 k\Omega$). For matching conditions ($Data = 0$ and $Search = 0$ (Fig.~\ref{fig:ExactTCAM}(a)), $Data = 1$ and $Search = 1$ (Fig.~\ref{fig:ExactTCAM}(e))), the FeSQUID connected in series with the superconducting hTron switches to its resistive state. As a result, the RWL current flows through this branch and we get a non-zero voltage drop in the ML (shown in Fig.~\ref{fig:ExactTCAM}(g)). On the other hand, for the mismatching conditions ($Data = 0$ and $Search = 1$ (Fig.~\ref{fig:ExactTCAM}(b)), $Data = 1$ and $Search = 0$ (Fig.~\ref{fig:ExactTCAM}(d))), the FeSQUID connected in series with the superconducting hTron remains in its superconducting state. As a result, we get zero voltage drop in the ML (shown in Fig.~\ref{fig:ExactTCAM}(g)). 

Now, for the 'd' case in the input search, we apply non-zero currents to both RBLs. As a result, in these cases shown in Fig.~\ref{fig:ExactTCAM}(c) and (f), both the hTrons switch to their resistive states which leads to the division of the applied RWL current between two branches. The current that flows through each of the branches ($I_{RWL}/2$) is not sufficient to switch the state of the FeSQUID to its resistive state. However, since both the hTrons become resistive, we get a non-zero voltage drop in the ML(shown in Fig.~\ref{fig:ExactTCAM}(g)), irrespective of the stored data in the FeSQUIDs. Fig.~\ref{fig:ExactTCAM}(g) shows value of ML voltages, power consumption, and energy consumption for all the possible combinations of a 1-bit TCAM.  

\begin{figure*}
\begin{center}
\includegraphics[width=\textwidth]{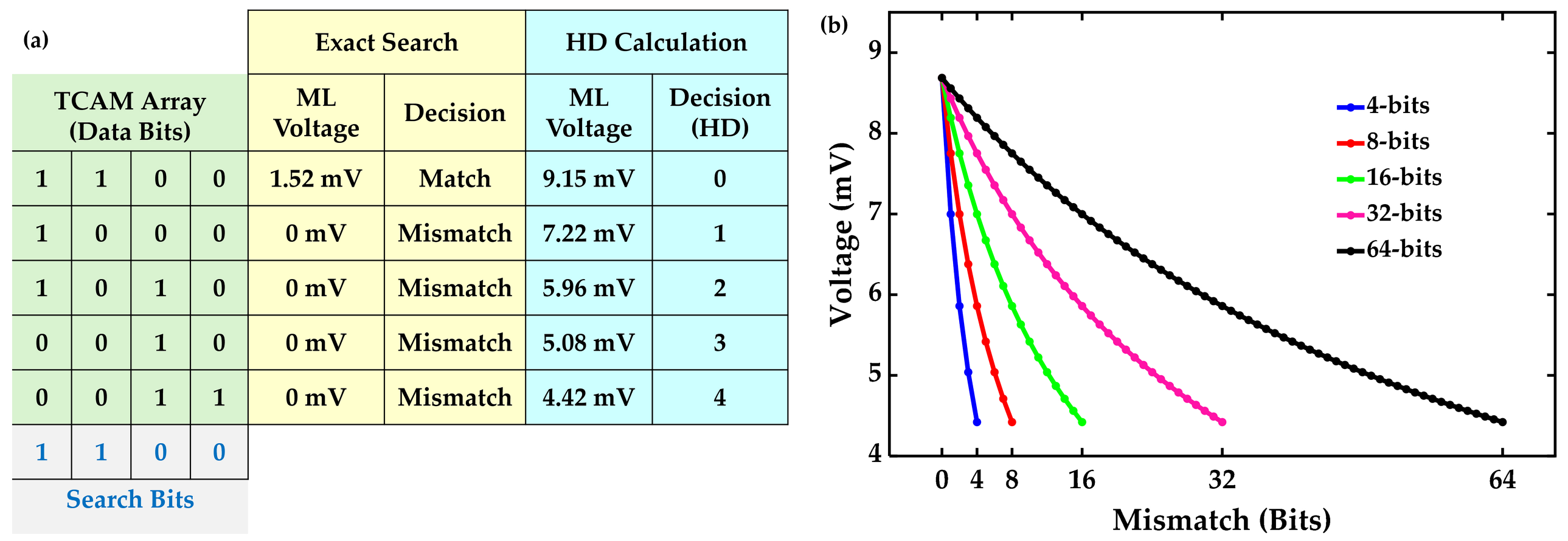}
\caption{ML voltage levels obtained for different number of data lengths and different amounts of mismatch between the input search and stored data.}
\label{fig:HD_TCAM}
\end{center}
\end{figure*}

Now, we discuss another capability of our proposed TCAM, which is the HD calculation. Fig.~\ref{fig:HD_TCAM}(a) shows a 4-bit TCAM where five different combinations of 4-bit data are stored in the TCAM array and one specific input search data is applied. As explained earlier in this section and Fig.~\ref{fig:ExactTCAM}, in the exact search mode, due to the choice of RWl current, we will get zero and non-zero ML voltages for any mismatch and complete match, respectively. However, in the HD calculation mode, we select the RWL current in a way so that the SQUID always gets a current that is larger than its both critical currents ($I_{C,Low}$ and $I_{C,High}$). Therefore, the SQUID never remains in the superconducting state. For the two states, we get two different values of resistance, as shown in Fig.~\ref{fig:Devices}(c). We design and choose the operating modes of the proposed TCAM in a way so that we get the highest ML voltage for the complete matching condition and the voltage drops with the increase in the mismatch. Fig.~\ref{fig:HD_TCAM}(a) shows the ML voltages for this mode for different amounts of mismatch in the 4-bit TCAM. We also extend our simulation to larger TCAM arrays and the values of ML voltage for different amounts of mismatch in different sizes of TCAM array in Fig.~\ref{fig:HD_TCAM}(b). The ML voltage for n-bit TCAM can be calculated using the following equation:
\begin{equation} 
\label{eq:vml}
V_{ML} = \left(n_{bits}\times I_{RWL}\right) \frac{1}{n_{bits} \times \frac{1}{50 k\Omega}+n_{match} \times \frac{1}{1.8 k\Omega}+\left(n_{bits} - n_{match}\right) \times \frac{1}{0.9 k\Omega}}
\end{equation}
where, $n_{bits}$ and $n_{match}$ represent the size of TCAM cells in a row and the number of matching bits in that row, respectively. To decode the HD from the ML voltage levels, any cryogenic voltage-based comparator and sense amplifier can be utilized \cite{alam2022superconducting}.  

\section{System-level Benchmarking}\label{sec4}

\begin{figure*}
    \centering
    \hfill
    \includegraphics[width=\textwidth]{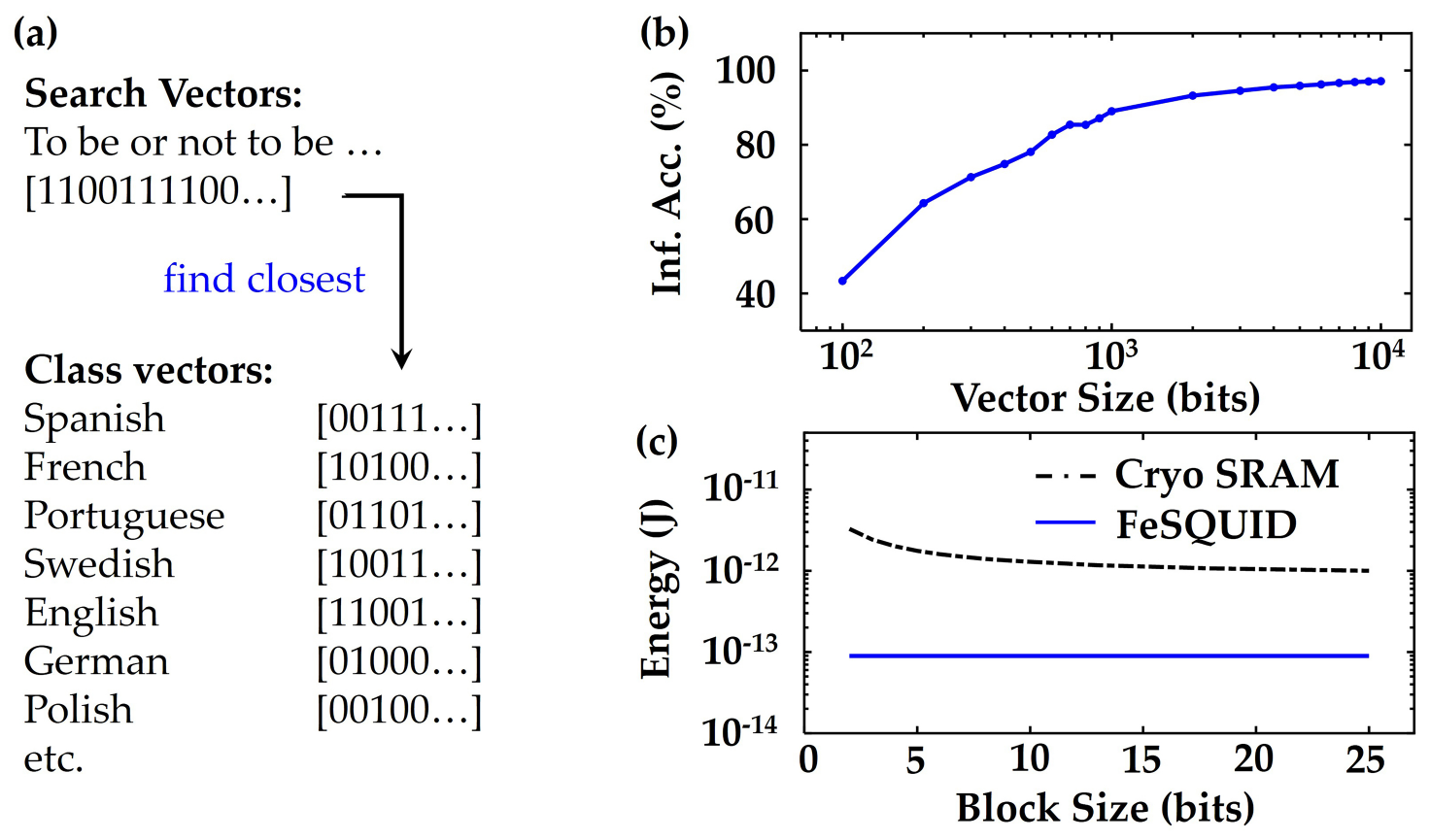}
    \hfill
    \caption{(a) Illustration of the \acrshort{hdc} algorithm for language recognition.
    The encoded search vector of the sample text that needs to be recognized is compared to all pre-trained class vectors.
    The most similar (i.e., closest) vector is then inferred as the result, and the respective class label is used to answer the search.
    (b) The inference accuracy of language recognition over the vector size.
    (c) Energy consumption of a single vector comparison with \qty{10000}{\bit} using \fesquid-based and \sram-based TCAM cells.
    Using the discussed energy estimation scheme, the \fesquid energy is constant with respect to block size.
    }
    \label{fig:hdc_application}
\end{figure*}

\begin{comment}

\begin{figure*}
    \centering
    \hfill
    \begin{subfigure}[c]{0.4\linewidth}
        \centering
        \includegraphics[height=6cm]{figures/language_recognition.pdf}
        \phantomsubcaption
        \label{fig:language_recognition}
    \end{subfigure}
    \hfill
    \begin{subfigure}[c]{0.55\linewidth}
        \centering
        \gnuplotloadfile[terminal=cairolatex, terminaloptions = {pdf size \convertlen{\linewidth}, 6cm}]{plot_files/energy_comparision.txt}
        \phantomsubcaption
        \label{fig:inference_accuracy}
        \phantomsubcaption
        \label{fig:sram_squid_energy}
    \end{subfigure}
    \hfill
    \caption{(a) Illustration of the \acrshort{hdc} algorithm for language recognition.
    The encoded search vector of the sample text that needs to be recognized is compared to all pre-trained class vectors.
    The most similar (i.e., closest) vector is then inferred as the result, and the respective class label is used to answer the search.
    (b) The inference accuracy of language recognition over the vector size.
    (c) Energy consumption of a single vector comparison with \qty{10000}{\bit} using \fesquid-based and \sram-based TCAM cells.
    Using the discussed energy estimation scheme, the \fesquid energy is constant with respect to block size.
    }
    \label{fig:hdc_application}
\end{figure*}
\end{comment}
With the ever-growing interest in machine learning, the size and data requirements of today's models are enormous.
However, many application scenarios have strict constraints for the system, e.g., in cryogenic environments, embedded systems, or IoT devices.
Here, efficient and computationally lightweight algorithms are key.
For these reasons, researchers have explored techniques to simplify models by quantizing conventional machine-learning approaches.
While reducing the model size, the fundamental requirements of large datasets and large training efforts are unchanged.

For these reasons, an emerging brain-inspired machine-learning concept called \hdc has been proposed and gained rapid interest.
In \hdc, information is stored as patterns in large vectors with thousands of elements.
By utilizing randomness and similarity, \hdc is inherently resilient to noisy data and errors in computing operations.
This enables the utilization of emerging hardware, which often may suffer from imprecise operation, e.g., due to variability.
\hdc has been showcased in various classical machine-learning tasks such as language recognition \cite{rahimi_robust_2016} (illustrated in \cref{fig:hdc_application}(a)), image classification \cite{holographic2017kleyko}, or wafer map defect pattern classification \cite{waferITC2021genssler}.

The data type of the vector elements can be selected, and real numbers, integer numbers, bipolar values, or binary values can be used.
Using binary makes \hdc very efficient as the operations can be implemented using basic logic gates.
For the similarity metric used during inference, the Hamming distance needs to be calculated, which is a combination of bit-wise XOR popcount.
This can be implemented using TCAM arrays that store the class vector bits and are searched with the bits of the query vector \cite{thomann2023tcam_hdc}.

The used vector size affects the precision of the final model, which is shown in \cref{fig:hdc_application}(b).
For instance, training the language recognition model with \num{10000} bits yields an inference accuracy of \qty{97.1}{\percent} while using half the size yields \qty{95.9}{\percent}.
Estimating the used energy consumption of a single vector comparison can be done using the following equation:
\begin{equation}
    E = \nbits \times \qty{5}{\micro\ampere} \times \vml \times \qty{0.3}{ns}
\end{equation}
where \nbits is the size of the vectors and \vml is the match line voltage depending on \nbits and \nmatch (see \cref{eq:vml}). Here, we consider the switching time of hTron as the only delay during the search operation in the TCAM cell. We use 0.3 ns for the superconducting to resistive switching of hTron devices, as reported in \cite{mccaughan2019superconducting}.
As most of the vectors in \hdc are orthogonal (i.e., a similarity of \qty{\approx 50}{\percent}), we assume the number of matching bits to be half of the total number of bits.
This gives a energy consumption of \qty{89.4}{\femto\joule} per vector comparison with \num{10000} bits, and \qty{44.7}{\femto\joule} with \num{5000} bits.

To put the numbers in perspective, we have implemented a block of \tcam cells using conventional \sram technology with \qty{5}{nm} \finfet transistors.
Similarly to the cryo \fesquid, we operate the circuits at \qty{4}{\kelvin}.
By cutting the large vectors into smaller sections, blocks are formed that calculate the Hamming distance in the analog domain.
Hence, the results in \cref{fig:hdc_application}(c) show a decreasing energy consumption for the \sram-based cells.
As fewer blocks are needed when increasing the block size, the comparison of an entire vector becomes more efficient.
In our energy estimation scheme for the cryo \fesquid, the energy depends on \vml, which is only subject to the ratio between block size to the number of mismatches, i.e., fixed to \qty{50}{\percent} here.
Still, the energy consumption of cryo \fesquid is over one order of magnitude smaller than cryo \sram.
For instance, using vectors with \qty{10000}{\bit} and a block size of \qty{10}{\bit}, \sram consumes \qty{1.29e-12}{\joule} compared to the \qty{8.94E-14}{\joule} of \fesquid.

\section{Methods}\label{sec11}

\subsection{Device Modeling and TCAM Simulation} \label{subsec51}

To perform the simulation of the proposed TCAM, we utilize Verilog-A-based compact models for FeSQUID and hTron from our previous works. To model the voltage-controlled switching of ferroelectric polarization, we use the Preisach model, which is calibrated with the experimental data reported in \cite{suleiman2021nonvolatile}. Then, we use equations (1) and (2) to model the effects of ferroelectric polarization on the superconducting energy gap and critical current of SQUID. Finally, we utilize the resistively and capacitively shunted junction (RCSJ) model to implement the current-voltage characteristics of SQUID, depending on the two polarization states of the ferroelectric material. 

For hTrons, we use a phenomenological Verilog-A model, which, at every timestep, compares the applied gate current and channel current with the switching thresholds. Based on the comparison, the model determines whether the channel will remain superconducting or switch to its resistive state. The switching thresholds and other device parameters were taken from the experimental data, reported in \cite{alam2023reconfigurable}. 

Using these two models, we developed a HSPICE framework to simulate and verify the TCAM design. The results mentioned in Fig.~\ref{fig:ExactTCAM} and~\ref{fig:HD_TCAM} are obtained using this simulation framework. 

\subsection{System-level Simulation}
\label{sec:sram_simulation}

The implementation of the \hdc algorithm for language recognition and \sram-based \tcam blocks is from a previous work \cite{thomann2023tcam_hdc}.
This allows us to change various parameters such as vector dimension, block size, or supply voltage.
The employed supply voltage for the \sram circuit simulations is \qty{0.7}{V}.
To simulate the transistor at these low temperatures accurately, we have employed a custom cryogenic compact model based on BSIM-CMG.
The respective model cards for NMOS and PMOS have been carefully calibrated to reproduce cryogenic measurements of mature \qty{5}{nm} \finfet technology \cite{parihar2023cryo_finfet}.

% supply voltage

% \textcolor{red}{Simon, please write this section for your part.}

\section{Conclusion}\label{sec13}

In this paper, we presented a cryogenic TCAM architecture leveraging the unique properties of FeSQUID technology. The proposed TCAM addresses key challenges faced by current cryogenic computing systems, such as the need for ultra-low power consumption, high-speed operation, and scalability. By combining the non-volatility of ferroelectric materials with the superconducting efficiency of SQUIDs, our design achieves substantial energy savings, with an average energy consumption of 1.36 aJ and 26.5 aJ for 1-bit binary and ternary searches, respectively. In addition to supporting exact match search operations, our TCAM enables Hamming distance (HD) calculation, which opens up new possibilities in error detection and correction, quantum error correction decoding, approximate search algorithms, and AI-based classification tasks. We demonstrate the advantages of FeSQUID-based TCAM array by benchmarking against TCAM with 5 nm FinFET-based cryogenic SRAM array for HDC-based language recognition task. The proposed FeSQUID-based TCAM needs over one order of magnitude lower energy. The parallel search capability and efficient data processing of this energy-efficient TCAM make it highly applicable to a range of cryogenic applications, including large-scale quantum computing, high-performance computing, space exploration, and data-intensive tasks in data centers.

\backmatter

%%\bmhead{Acknowledgements}

%Acknowledgements are not compulsory. Where included they should be brief. Grant or contribution numbers may be acknowledged.

%Please refer to Journal-level guidance for any specific requirements.

\section*{Declarations}

\begin{itemize}

\item \textbf{Funding:} This work was supported in part by the Department of Energy (DOE) under award no. DE-SC0024328.
S. A. was supported with funds provided by the Science Alliance, a Tennessee Higher Education Commission center of excellence administered by The University of Tennessee-Oak Ridge Innovation Institute on behalf of The University of Tennessee, Knoxville.

\item \textbf{Conflict of interest/Competing interests:} The authors declare no conflict of interest.

%\item Ethics approval and consent to participate
%\item Consent for publication
\item \textbf{Data availability:} The data that support the plots within this paper and other findings of this study are available from the corresponding author upon reasonable request.

\item \textbf{Author contribution:} S.A. conceived the idea, designed the cryogenic TCAM, developed the simulation framework, and performed the simulations to verify the design. S.T. and S.S.P. performed the system-level simulations and benchmarking. S.A. and S.T. prepared the draft manuscript. All the authors commented on the results, reviewed the manuscript, and contributed to writing the manuscript. A.A., H.A., and Y.S.C. supervised the project. 

\end{itemize}

\bibliography{sn-bibliography}% common bib file

\begin{thebibliography}{10}
\providecommand{\url}[1]{{#1}}
\providecommand{\urlprefix}{URL }
\providecommand{\doi}[1]{\url{https://doi.org/#1}}
\bibcommenthead

\bibitem{alam2023cryogenic}
S.~Alam, M.S. Hossain, S.R. Srinivasa, A.~Aziz, Cryogenic memory technologies.
\newblock Nature Electronics \textbf{6}(3), 185--198 (2023)

\bibitem{hornibrook2015cryogenic}
J.~Hornibrook, J.~Colless, I.C. Lamb, S.~Pauka, H.~Lu, A.~Gossard, J.~Watson, G.~Gardner, S.~Fallahi, M.~Manfra, et~al., Cryogenic control architecture for large-scale quantum computing.
\newblock Physical Review Applied \textbf{3}(2), 024010 (2015)

\bibitem{alam2022cryocim}
S.~Alam, M.M. Islam, M.S. Hossain, A.~Jaiswal, A.~Aziz, Cryocim: Cryogenic compute-in-memory based on the quantum anomalous hall effect.
\newblock Applied Physics Letters \textbf{120}(14) (2022)

\bibitem{holmes2013energy}
D.S. Holmes, A.L. Ripple, M.A. Manheimer, Energy-efficient superconducting computing—power budgets and requirements.
\newblock IEEE Transactions on Applied Superconductivity \textbf{23}(3), 1701610--1701610 (2013)

\bibitem{huang2022survey}
J.~Huang, R.~Fu, X.~Ye, D.~Fan, A survey on superconducting computing technology: circuits, architectures and design tools.
\newblock CCF Transactions on High Performance Computing \textbf{4}(1), 1--22 (2022)

\bibitem{das2017large}
R.N. Das, V.~Bolkhovsky, S.K. Tolpygo, P.~Gouker, L.M. Johnson, E.A. Dauler, M.A. Gouker, \emph{Large scale cryogenic integration approach for superconducting high-performance computing}, in \emph{2017 IEEE 67th Electronic Components and Technology Conference (ECTC)} (IEEE, 2017), pp. 675--683

\bibitem{alam2021non}
S.~Alam, M.S. Hossain, A.~Aziz, A non-volatile cryogenic random-access memory based on the quantum anomalous hall effect.
\newblock Scientific Reports \textbf{11}(1), 7892 (2021)

\bibitem{ghoshal1993superconductor}
U.~Ghoshal, H.~Kroger, T.~Van~Duzer, Superconductor-semiconductor memories.
\newblock IEEE transactions on applied superconductivity \textbf{3}(1), 2315--2318 (1993)

\bibitem{alam2022cryogenic}
S.~Alam, M.M. Islam, M.S. Hossain, K.~Ni, V.~Narayanan, A.~Aziz, \emph{Cryogenic memory array based on ferroelectric squid and heater cryotron}, in \emph{2022 Device Research Conference (DRC)} (IEEE, 2022), pp. 1--2

\bibitem{alam2021cryogenic}
S.~Alam, M.S. Hossain, A.~Aziz, A cryogenic memory array based on superconducting memristors.
\newblock Applied Physics Letters \textbf{119}(8) (2021)

\bibitem{tannu2017cryogenic}
S.S. Tannu, D.M. Carmean, M.K. Qureshi, \emph{Cryogenic-DRAM based memory system for scalable quantum computers: A feasibility study}, in \emph{Proceedings of the International Symposium on Memory Systems} (2017), pp. 189--195

\bibitem{islam2023quantum}
M.M. Islam, J.~Hutchins, S.~Alam, M.S. Hossain, A.~Jaiswal, A.~Aziz, \emph{Quantum Anomalous Hall Effect-Based Variation Robust Binary Content Addressable Memory}, in \emph{2023 IEEE 66th International Midwest Symposium on Circuits and Systems (MWSCAS)} (IEEE, 2023), pp. 331--335

\bibitem{ni2019ferroelectric}
K.~Ni, X.~Yin, A.F. Laguna, S.~Joshi, S.~D{\"u}nkel, M.~Trentzsch, J.~M{\"u}ller, S.~Beyer, M.~Niemier, X.S. Hu, et~al., Ferroelectric ternary content-addressable memory for one-shot learning.
\newblock Nature Electronics \textbf{2}(11), 521--529 (2019)

\bibitem{suleiman2021nonvolatile}
M.~Suleiman, M.F. Sarott, M.~Trassin, M.~Badarne, Y.~Ivry, Nonvolatile voltage-tunable ferroelectric-superconducting quantum interference memory devices.
\newblock Applied Physics Letters \textbf{119}(11) (2021)

\bibitem{huang2021direct}
Q.~Huang, Z.~Chen, M.J. Cabral, F.~Wang, S.~Zhang, F.~Li, Y.~Li, S.P. Ringer, H.~Luo, Y.W. Mai, et~al., Direct observation of nanoscale dynamics of ferroelectric degradation.
\newblock Nature communications \textbf{12}(1), 2095 (2021)

\bibitem{crassous2011nanoscale}
A.~Crassous, R.~Bernard, S.~Fusil, K.~Bouzehouane, D.~Le~Bourdais, S.~Enouz-Vedrenne, J.~Briatico, M.~Bibes, A.~Barth{\'e}l{\'e}my, J.E. Villegas, Nanoscale electrostatic manipulation of magnetic flux quanta in ferroelectric/superconductor bifeo 3/yba 2 cu 3 o 7- $\delta$ heterostructures.
\newblock Physical review letters \textbf{107}(24), 247002 (2011)

\bibitem{bardeen1957theory}
J.~Bardeen, L.N. Cooper, J.R. Schrieffer, Theory of superconductivity.
\newblock Physical review \textbf{108}(5), 1175 (1957)

\bibitem{alam2020compact}
S.~Alam, M.A. Jahangir, A.~Aziz, A compact model for superconductor-insulator-superconductor (sis) josephson junctions.
\newblock IEEE Electron Device Letters \textbf{41}(8), 1249--1252 (2020)

\bibitem{ambegaokar1963tunneling}
V.~Ambegaokar, A.~Baratoff, Tunneling between superconductors.
\newblock Physical Review Letters \textbf{10}(11), 486 (1963)

\bibitem{alam2024voltage}
S.~Alam, M.S. Hossain, K.~Ni, V.~Narayanan, A.~Aziz, Voltage-controlled cryogenic boolean logic gates based on ferroelectric squid and heater cryotron.
\newblock Journal of Applied Physics \textbf{135}(1) (2024)

\bibitem{alam2023Dac}
S.~Alam, J.~Hutchins, M.S. Hossain, K.~Ni, V.~Narayanan, A.~Aziz, \emph{Cryogenic in-memory matrix-vector multiplication using ferroelectric superconducting quantum interference device (FE-SQUID)}, in \emph{2023 60th ACM/IEEE Design Automation Conference (DAC)} (IEEE, 2023), pp. 1--6

\bibitem{mccaughan2014superconducting}
A.N. McCaughan, K.K. Berggren, A superconducting-nanowire three-terminal electrothermal device.
\newblock Nano letters \textbf{14}(10), 5748--5753 (2014)

\bibitem{mccaughan2019superconducting}
A.N. McCaughan, V.B. Verma, S.M. Buckley, J.~Allmaras, A.~Kozorezov, A.~Tait, S.~Nam, J.~Shainline, A superconducting thermal switch with ultrahigh impedance for interfacing superconductors to semiconductors.
\newblock Nature electronics \textbf{2}(10), 451--456 (2019)

\bibitem{nguyen2020cryogenic}
M.H. Nguyen, G.J. Ribeill, M.V. Gustafsson, S.~Shi, S.V. Aradhya, A.P. Wagner, L.M. Ranzani, L.~Zhu, R.~Baghdadi, B.~Butters, et~al., Cryogenic memory architecture integrating spin hall effect based magnetic memory and superconductive cryotron devices.
\newblock Scientific reports \textbf{10}(1), 248 (2020)

\bibitem{alam2023reconfigurable}
S.~Alam, D.S. Rampini, B.G. Oripov, A.N. McCaughan, A.~Aziz, Cryogenic reconfigurable logic with superconducting heater cryotron: Enhancing area efficiency and enabling camouflaged processors.
\newblock Applied Physics Letters \textbf{123}(15) (2023)

\bibitem{alam2024ultra}
S.~Alam, A.~Aziz, \emph{Ultra-Area-Efficient Cryogenic XNOR Logic Gate with Superconducting Heater Cryotron to Advance High-Performance Computing}, in \emph{Proceedings of the Great Lakes Symposium on VLSI 2024} (2024), pp. 651--656

\bibitem{islam2023review}
M.M. Islam, S.~Alam, M.S. Hossain, K.~Roy, A.~Aziz, A review of cryogenic neuromorphic hardware.
\newblock Journal of Applied Physics \textbf{133}(7) (2023)

\bibitem{islam2022dynamically}
M.M. Islam, S.~Alam, M.S. Hossain, A.~Aziz, \emph{Dynamically reconfigurable cryogenic spiking neuron based on superconducting memristor}, in \emph{2022 IEEE 22nd International Conference on Nanotechnology (NANO)} (IEEE, 2022), pp. 307--310

\bibitem{islam2023synapse}
M.M. Islam, S.~Alam, M.R.I. Udoy, M.S. Hossain, A.~Aziz, \emph{A cryogenic artificial synapse based on superconducting memristor}, in \emph{Proceedings of the Great Lakes Symposium on VLSI 2023} (2023), pp. 143--148

\bibitem{alam2022superconducting}
S.~Alam, M.M. Islam, M.S. Hossain, A.~Aziz, \emph{Superconducting josephson junction fet-based cryogenic voltage sense amplifier}, in \emph{2022 Device Research Conference (DRC)} (IEEE, 2022), pp. 1--2

\bibitem{rahimi_robust_2016}
A.~Rahimi, P.~Kanerva, J.M. Rabaey, \emph{A {Robust} and {Energy}-{Efficient} {Classifier} {Using} {Brain}-{Inspired} {Hyperdimensional} {Computing}}, in \emph{Proceedings of the 2016 {International} {Symposium} on {Low} {Power} {Electronics} and {Design} - {ISLPED} '16} (ACM Press, San Francisco Airport, CA, USA, 2016), pp. 64--69.
\newblock \doi{10.1145/2934583.2934624}.
\newblock \urlprefix\url{http://dl.acm.org/citation.cfm?doid=2934583.2934624}

\bibitem{holographic2017kleyko}
D.~{Kleyko}, E.~{Osipov}, A.~{Senior}, A.I. {Khan}, Y.A. {Şekerciogğlu}, Holographic graph neuron: A bioinspired architecture for pattern processing.
\newblock IEEE Transactions on Neural Networks and Learning Systems \textbf{28}(6), 1250--1262 (2017).
\newblock \doi{10.1109/TNNLS.2016.2535338}

\bibitem{waferITC2021genssler}
P.R. Genssler, H.~Amrouch, \emph{Brain-Inspired Computing for Wafer Map Defect Pattern Classification}, in \emph{2021 IEEE International Test Conference (ITC)} (2021), pp. 123--132.
\newblock \doi{10.1109/ITC50571.2021.00020}

\bibitem{thomann2023tcam_hdc}
S.~Thomann, P.R. Genssler, H.~Amrouch, Hw/sw co-design for reliable tcam- based in-memory brain-inspired hyperdimensional computing.
\newblock IEEE Transactions on Computers \textbf{72}(8), 2404--2417 (2023).
\newblock \doi{10.1109/TC.2023.3248286}

\bibitem{parihar2023cryo_finfet}
S.S. Parihar, V.M. van Santen, S.~Thomann, G.~Pahwa, Y.S. Chauhan, H.~Amrouch, Cryogenic cmos for quantum processing: 5-nm finfet-based sram arrays at 10 k.
\newblock IEEE Transactions on Circuits and Systems I: Regular Papers \textbf{70}(8), 3089--3102 (2023).
\newblock \doi{10.1109/TCSI.2023.3278351}

\end{thebibliography}

\end{document}